\documentclass[11pt,a4paper]{article}
\usepackage{jcappub}

\usepackage[utf8]{inputenc} %Support for additional characters in bibliography
\usepackage{amsfonts,amssymb, mathrsfs, amsmath, esint,bm,enumitem}
\allowdisplaybreaks

\usepackage{subcaption}
\usepackage{ragged2e}
\DeclareCaptionJustification{justified}{\justifying}
\usepackage{diagbox}
\usepackage{latexsym}
\usepackage{graphicx}
\usepackage[dvipsnames]{xcolor}
\usepackage{booktabs,multirow}
\usepackage{titlesec} % to change style of \paragraph{} header
\usepackage{physics}
\usepackage[normalem]{ulem}

\usepackage{tikz}
\usetikzlibrary{positioning,calc}

\usepackage{hyperref}
\hypersetup{colorlinks, citecolor=ForestGreen, linkcolor=BrickRed, urlcolor=RedViolet}

\usepackage[capitalize]{cleveref}
\usepackage{float}

\newcommand{\out}[1]{\textcolor{OliveGreen}{\sout{#1}}}
\renewcommand{\out}[1]{}

\newcommand{\plik}{\texttt{Plik}\,}
\newcommand{\commander}{\texttt{Commander}\,}
\newcommand{\simall}{\texttt{SimAll}\,}
\newcommand{\npipe}{\texttt{NPIPE}\,}
\newcommand{\hillipop}{\texttt{HiLLiPoP}\,}
\newcommand{\lollipop}{\texttt{LoLLiPoP}\,}
\newcommand{\camspec}{\texttt{CamSpec}\,}

\newcommand{\planck}{\textbf{P18}}
\newcommand{\planckH}{\textbf{P20$_{\rm \bf H}$}}
\newcommand{\planckC}{\textbf{P20$_{\rm \bf C}$}}
\newcommand{\desi}{$\mathbf{}$\textbf{DESI}}

\newcommand{\eboss}{$\mathbf{}$\textbf{SDSS}$_\mathbf{16}$}
\newcommand{\desisdss}{$\mathbf{}$\textbf{DESI/SDSS}}
\newcommand{\baoplus}{$\mathbf{}$\textbf{SDSS}$_\mathbf{f\sigma_8}$}
\newcommand{\pantheon}{$\mathbf{}${\bf Pantheon}}
\newcommand{\DES}{$\mathbf{}${\bf DES-SN}}

\newcommand{\shoes}{$\mathbf{ H_0}$}

\newcommand{\DNeff}{\Delta N_\text{eff}}

\newcommand{\beq}{\begin{equation}}
\newcommand{\eeq}{\end{equation}}

\title{Neutrino mass bounds from DESI 2024 are relaxed 
by Planck PR4
\\and cosmological supernovae\\
}

\author[a]{Itamar J. Allali}
\author[b,c]{Alessio Notari}

\affiliation[a]{ Department of Physics, Brown University, Providence, RI 02912, USA\looseness=-1}
\affiliation[b]{Departament de F\'isica Qu\`antica i Astrofis\'ica \& Institut de Ci\`encies del Cosmos (ICCUB), Universitat de Barcelona, Mart\'i i Franqu\`es 1, 08028 Barcelona, Spain. 		\looseness=-1}
 \affiliation[c]{ Galileo Galilei Institute for theoretical physics, Centro Nazionale INFN di Studi Avanzati Largo Enrico Fermi 2, I-50125, Firenze, Italy		\looseness=-1}

\emailAdd{itamar\_allali@brown.edu}
\emailAdd{notari@fqa.ub.edu}

\abstract{
The recent DESI 2024 Baryon Acoustic Oscillations (BAO) measurements  combined with the CMB data from the Planck 18 PR3 dataset and the Planck PR4+ACT DR6 lensing data, with a prior on the sum of the neutrino masses $\sum m_\nu>0$, leads to a strong constraint, $\sum m_\nu<0.072$ eV, which would exclude the inverted neutrino hierarchy and put some tension on even the standard hierarchy.  We show that actually this bound gets significantly relaxed when combining the new DESI measurements with the \hillipop + \lollipop~likelihoods, based on the Planck 2020 PR4 dataset, and with supernovae datasets. We note that the fact that neutrino masses are pushed towards zero, and even towards negative values, is known to be correlated with the so-called  $A_L$ tension, a mismatch between lensing and power spectrum measurements in the Planck PR3 data, which is reduced by \hillipop + \lollipop~to less than 1$\sigma$. We find $\sum m_\nu<0.1$ eV and  $\sum m_\nu<0.12$ eV, with the supernovae \emph{Pantheon+} and \emph{DES-SN5YR} datasets respectively. The shift caused by these datasets 
is more compatible with the expectations from neutrino oscillation experiments, and both the normal and inverted hierarchy scenarios remain now viable, even with the $\sum m_\nu>0$  prior. Finally, we analyze neutrino mass bounds in an extension of $\Lambda$CDM that addresses the $H_0$ tension, with extra fluid Dark Radiation, finding that in such models bounds are further relaxed and 
the posterior probability for $\sum m_\nu$ begins to exhibit a peak at positive values.
}

\begin{document}
\maketitle
\vspace{1em}\noindent

%%%%%%%%%%%%%%%%%%%%%%%%%%%%%%%%%%%%%%%%%%%%%%%%%%%%%%%%%%%%%%%%%%%%%%%%
\section{Introduction}
\label{sec:introduction}

Cosmological observations are at present the most promising way to detect for the first time the sum of neutrino masses. Nonetheless, the recent combination of datasets presented by  the Dark Energy Spectroscopic Instrument (DESI) collaboration~\cite{DESI:2024mwx}, including their new data release on Baryon Acoustic Oscillations (BAO) together with the Planck CMB 2018 data~\cite{Planck:2018vyg} (the \plik, \commander, and \simall likelihoods based on the 2018 PR3 dataset on Temperature and Polarization, together with the \npipe~PR4 Planck CMB lensing reconstruction~\cite{Carron:2022eyg} and the lensing data from the Data Release 6 of the Atacama Cosmology Telescope~\cite{ACT:2023dou}), is showing only a (quite stringent) upper bound on the sum of neutrino masses, $\sum m_\nu<0.072$~eV~\cite{DESI:2024mwx}, when imposing the most conservative prior $\sum m_\nu>0$. There is no hint of a nonzero mass and the posterior probability actually shows a cusp at zero, so that the peak of the distribution, if extended with a Gaussian~\cite{eBOSS:2020yzd}, would even go to (unphysical) negative values (see also~\cite{Craig:2024tky}). Since positive neutrino masses imply a suppression of the matter power spectrum, this would mean that such data prefer an enhancement of the spectrum.

However, it is known that this preference in the direction of negative values is correlated to the lensing ``anomaly," or tension, present in the likelihoods based on Planck 2018 data~\cite{Planck:2018vyg}, i.e the fact that the ad hoc parameter $A_L$, that rescales the deflection power spectrum used to lens the primordial CMB power spectra, is larger than 1 when it is left free to vary, instead of being consistent with its real value $A_L=1$. Forcing $A_L=1$ pushes instead the neutrino masses towards negative values~\cite{Planck:2018vyg}. Recently, new likelihoods for the final (PR4) Planck CMB data release have been published~\cite{Tristram:2023haj}, both for high-$\ell$ TT, TE and EE spectra (\hillipop) and for the low-$\ell$ EE polarization spectra (\lollipop), to be used together with the Planck18 low-$\ell$ TT data. Such new likelihoods have been shown to lead to $A_L= 1.039 \pm 0.052 $ in $\Lambda$CDM, consistent with the expected value of unity. It has been already shown using CMB data alone~\cite{Tristram:2023haj}, that as a result of this shift, the neutrino masses move to more positive values.

The aim of this work is to assess the status of the preference for positive neutrino masses employing such new CMB likelihoods, combined together with the new released  BAO data from galaxies and quasars~\cite{DESI:2024uvr} at redshifts $0.3 \lesssim z\lesssim 1.5$ and from the Lyman-$\alpha$ forest~\cite{DESI:2024lzq} by DESI~\cite{DESI:2024mwx}  2024. We will also check the impact on neutrino masses of Supernovae datasets, i.e. Pantheon$+$~\cite{Scolnic:2021amr} and DES-SN5YR~\cite{DES:2024tys}.

We will analyse such bounds in the context of the $\Lambda$CDM model, with varying neutrino masses. Subsequently, we will also consider neutrino mass bounds in extensions of the $\Lambda$CDM model that have been recently proposed to address the Hubble tension with the addition of a Dark Radiation (DR) component~\cite{Allali:2024cji}.

In all the analyses, we will apply a prior $\sum m_\nu>0$, i.e.~we assume here no prior information from  neutrino oscillation experiments, in order to have a fully independent measurement of neutrino masses. 

\section{Models and datasets}\label{sec:model}

We will first  study  the simple $\Lambda$CDM spatially-flat cosmological model with free sum of neutrino masses, the $\Lambda$CDM+$\sum m_\nu$ model. We assume for simplicity the three neutrinos to have  the same mass, since it has been shown that  current experiments are sensitive only to the sum of neutrino masses, irrespective of how are they distributed~\cite{CORE:2016npo}.

We perform a Bayesian analysis using {\tt CLASS}~\cite{Lesgourgues:2011re, Blas:2011rf} to solve for the cosmological evolution and either {\tt MontePython}~\cite{Audren:2012wb, Brinckmann:2018cvx} or {\tt Cobaya}~\cite{Torrado:2020dgo,2019ascl.soft10019T}
to collect Markov Chain Monte Carlo (MCMC) samples. We obtain posteriors and figures using {\tt GetDist}~\cite{Lewis:2019xzd}. We consider various combinations of datasets, as follows.

In~\cref{sec:planck} we will explore three different Planck likelihoods for CMB data:

\begin{itemize}

\item{\planck: the Planck 2018 high-$\ell$ TT, TE, EE \plik, low-$\ell$ TT \commander, and low-$\ell$ EE \simall likelihoods, together with the Planck 2018 lensing data~\cite{Planck:2019nip}};

\item{\planckH: the  \hillipop+ \lollipop likelihoods~\cite{Tristram:2023haj}, based on the final Planck data release (PR4)~\cite{Planck:2020olo}. In particular: the \hillipop likelihood at high-$\ell$ for TT, TE, EE;  the \lollipop likelihood  for low-$\ell$  EE; the Planck 2018 \commander likelihood for low-$\ell$ TT and Planck 2018 CMB lensing data~\cite{Planck:2019nip}};

\item{\planckC: The \camspec likelihood~\cite{Efstathiou:2019mdh}, updated by~\cite{Rosenberg:2022sdy} to the 2020 Planck PR4 data release~\cite{Planck:2020olo}, at high-$\ell$ for TT, TE, EE; the Planck 2018 \commander and \simall likelihoods for low$\ell$ TT and EE, respectively, and the Planck 2020 PR4 lensing likelihood \cite{Carron:2022eyg}}.

\end{itemize}

Then, in~\cref{sec:SN}, we will explore the effects of including different sets of cosmological supernovae:

\begin{itemize}
\item{\pantheon: The Pantheon+ supernovae compilation~\cite{Scolnic:2021amr}.}

\item{\DES:  The DES-SN5YR supernovae compilation~\cite{DES:2024tys}.}

\end{itemize}

For most of this work, we will focus on the BAO measurements from DESI. In~\cref{sec:BAO}, we will also compare to other BAO measurements. The BAO datasets under consideration are:
\begin{itemize}
\item{\desi: BAO measurements from DESI 2024~\cite{DESI:2024mwx} at effective redshifts $z=0.3$, $ 0.51$ $0.71$, $0.93$, $1.32$, $1.49$, $2.33$;  }

\item{\eboss: BAO measurements from 6dFGS at $z = 0.106$~\cite{Beutler:2011hx}; SDSS MGS at $z = 0.15$~\cite{Ross:2014qpa}; and SDSS eBOSS DR16 measurements \cite{eBOSS:2020yzd}, including DR12 galaxies~\cite{BOSS:2016wmc}, and DR16 LRG~\cite{eBOSS:2020lta,eBOSS:2020hur}, QSO~\cite{eBOSS:2020gbb,eBOSS:2020uxp}, ELG~\cite{eBOSS:2020qek,eBOSS:2020fvk}, Lyman-$\alpha$, and Lyman-$\alpha$ $\times$ QSO~\cite{eBOSS:2020tmo}.}

\item{\baoplus: The same BAO measurements given in \eboss, while using the full-shape likelihoods for LRG, ELG, QSO, Lyman-$\alpha$, and Lyman-$\alpha$ $\times$ QSO~\cite{eBOSS:2020yzd}, which include constraints on $f\sigma_8$ from redshift-space distortions.}

\item{\desisdss: BAO measurements from the combination suggested in~\cite{DESI:2024mwx} that merges DESI 2024 with previous SDSS measurements, choosing for each bin the measurement with the highest precision to date.}

\end{itemize}

After discussing the status of neutrino masses in the simplest setup, we will also extend our analysis to models beyond $\Lambda$CDM, that have recently been shown to address the so-called Hubble tension, i.e. the tension on the determination of the present Hubble rate from the above datasets with the following local direct measurement of the expansion rate by the SH$0$ES collaboration:

\begin{itemize}
\item{\shoes: the measurement of the intrinsic SNIa magnitude $M_b=-19.253\pm0.027$~\cite{Riess:2021jrx}, which uses a Cepheid-calibrated distance ladder.  We add this always in combination with Pantheon+ data, as implemented in the {\tt Pantheon\_Plus\_SHOES} likelihood in {\tt MontePython}.\footnote{An even newer measurement from the collaboration is given in~\cite{Breuval:2024lsv}, but we use the value in~\cite{Riess:2021jrx} because of the available combination with Pantheon+.}}
\end{itemize}

Such models extend $\Lambda$CDM by including a new Dark Radiation (DR) component, which lowers the tension~\cite{Allali:2024cji}, below 3$\sigma$ and as low as $1.8 \sigma$, depending on the specific realization (free-streaming or fluid DR, present before BBN or produced after BBN\footnote{We note that constraints from primordial element abundances, which we do not include in this work, are not relevant when DR is produced after BBN.}) 
and on the combination of datasets. Given the lower degree of tension, we are allowed in this case to combine with the \shoes  \, measurement, interpreting the  tension as a moderate statistical fluctuation. In~\cref{sec:fluidDR}, we will focus on one particular choice for the DR, the fluid DR present before the epoch of BBN, for simplicity.

%%%%%%%%%%%%%%%%%%%%%%%%%%%%%%%%%%%%%%%%%%%%%%%%%%%%%%%%%%%%%%%%%%%%%%%

%%%%%%%%%%%%%%%%%%%%%%%%%%%%%%%%%%%%%%%%%%%%%%%%%%%%%%%%%%%%%%%%%%%%%%%%
\section{New constraints on neutrino masses in $\Lambda$CDM+$\sum m_\nu$} 
\label{sec:constraints}

In this section we  conservatively consider the $\Lambda$CDM model with variable neutrino masses, even if such model is: (1) in strong tension with SH$0$ES, and (2) mildly disfavoured compared to time varying dark-energy scenarios when not considering SH$0$ES (e.g. with respect to the so-called $w_0 w_a$CDM model~\cite{DESI:2024mwx}, which however points to the unphysical region with equation of state $w<-1$, or physically viable models, such as the ``ramp" quintessence model~\cite{Notari:2024rti}).

Within the $\Lambda$CDM$+\sum m_\nu$ model, the DESI collaboration finds~\cite{DESI:2024mwx} 
$\sum  m_\nu < 0.072$~eV  (95\%CL, using DESI and Planck 2018 TT, TE, EE likelihoods, with PR4+ACT DR6 lensing data),
which improves substantially on the analogous previous bound,
$\sum  m_\nu < 0.12$~eV  (95\%CL, from Planck 2018 combined with SDSS DR12 BAO~\cite{Planck:2018vyg}). 
At face value, this new bound excludes the inverted hierarchy case ($\sum m_\nu> 0.10$ eV) and starts to put some pressure even on the normal hierarchy case~\footnote{Note however that such a strong conclusion is not robust under the change of prior, i.e. it does not hold when using the prior based on neutrino oscillations, $\sum m_\nu>0.06$ eV.}. The situation, however, changes substantially when exploring various combinations of datasets, as discussed below. 

\begin{figure}[t]
    \centering
    \includegraphics[width=0.6\textwidth]{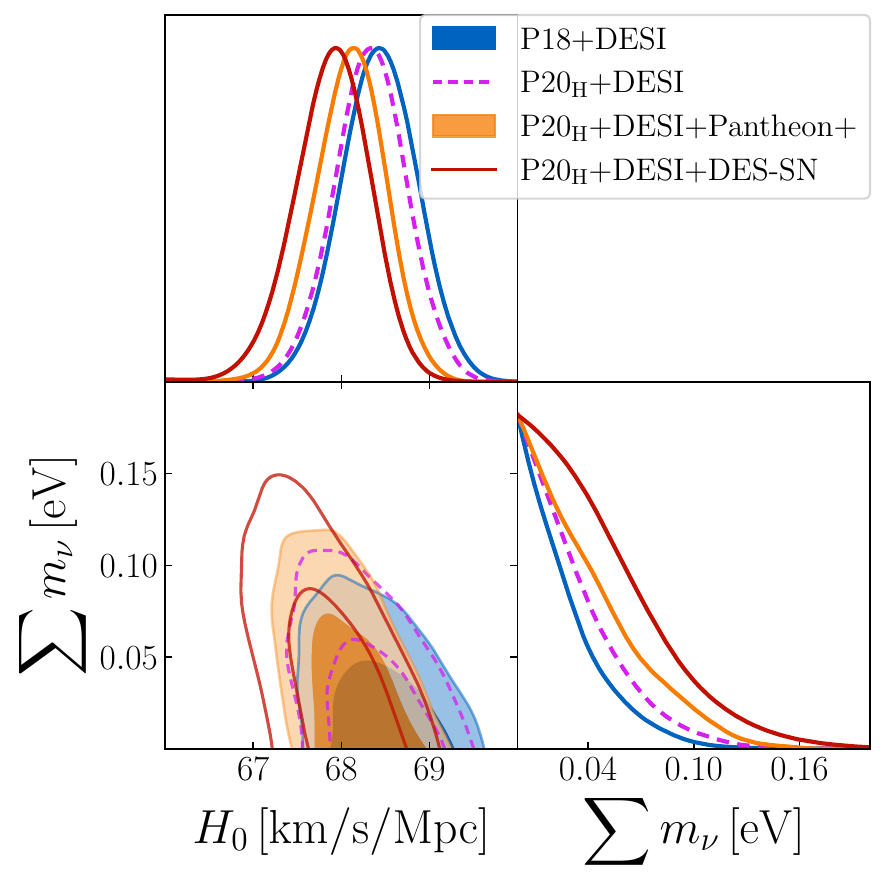}
    \caption{One- and two-dimensional posteriors for $H_0$ and $\sum m_\nu$ in the $\Lambda$CDM + $\sum m_\nu$ model, fitting to combinations of the Planck 2018 likelihoods with DESI BAO, compared to using the Planck 2020 \hillipop+\lollipop~likelihoods with DESI BAO and the sets of cosmological supernovae from Pantheon+ and DES-SN5YR.}
    \label{fig:p18_p20_desi_pantheon_DES}
\end{figure}

\subsection{Effects of Planck Likelihoods}\label{sec:planck}
  
First, we discuss the effect of including more recent Planck likelihoods, from the PR4 2020 data release.  The bound gets substantially relaxed by making use of the recent \hillipop+\\\lollipop (\planckH) likelihoods, leading to:
\begin{center}
$\sum m_\nu < 0.086~{\rm eV} $~(95\%CL,~\planckH+\desi).
\end{center}
The weakening of the bound compared to the case with \planck \, is consistent with the expectation that a smaller $A_L$ should lead to larger neutrino masses~\cite{Couchot:2017pvz, Planck:2019nip, Tristram:2023haj}. Note that we compare here to our \planck~combination which uses the Planck 2018 lensing, giving $\sum m_\nu < 0.073$ eV (see \cref{tab:lcdm_bounds}), rather than comparing to the constraint in~\cite{DESI:2024mwx} which uses a different lensing likelihood.

Using instead the \camspec likelihood which has also been updated to PR4, we find an intermediate result:
\begin{center}
$\sum m_\nu < 0.080~{\rm eV} $~(95\%CL,~\planckC+\desi).
\end{center}
These constraints, as well the constraint from the combination of data \planck~ defined above, are summarized in~\cref{tab:lcdm_bounds}. In addition, in~\cref{fig:p18_p20_desi_pantheon_DES}, one can compare the posteriors of $\sum m_\nu$ for \planck+\desi~ and \planckH+\desi, noting that the latter contour shows a relaxed constraint.

Since \planckH~ and \planckC~ use the most up-to-date set of data from Planck, and further, since it is has been shown that \hillipop+\lollipop have been the most effective at eliminating the $A_L$ problem in the Planck data, we will take the combination \planckH~ to be the preferred Planck dataset for the remainder of this work.

\subsection{Effects of Supernovae likelihoods}\label{sec:SN}

\begin{table}[t]
    \centering
    \begin{tabular}{|l|c||l|c|}
    \hline
         Dataset & $\sum m_\nu$ & Dataset & $\sum m_\nu$  \\
         \hline
         \planck+\desi & $< 0.073$  & 
         \planckC+\desi & $< 0.080$ \\
         ~~+~\pantheon & $< 0.086$ &
         \planckH+\eboss & $< 0.14$ \\
          ~~+~\DES &  $< 0.094$&
         \planckH+\baoplus & $< 0.11$  \\
       \planckH+\desi & $< 0.086$  & \planckH+\desisdss & $<0.11$ \\
         ~~+~\pantheon & $< 0.099$  & ~~+~\pantheon & $<0.12$ \\
          ~~+~\DES & $< 0.11$ &   ~~+~\DES  & $<0.13$\\
         \hline
    \end{tabular}
    \caption{ 95\%CL upper bounds on the neutrino mass sum for $\Lambda$CDM + $\sum m_\nu$ model, comparing fits to various datasets.}
    \label{tab:lcdm_bounds}
\end{table}

Adding supernovae, the bounds get further relaxed 
\begin{center}
$\sum m_\nu < 0.099~{\rm eV}$~(95\%CL,~\planckH+\desi+\pantheon),
\end{center}
\begin{center}
$\sum m_\nu < 0.11~{\rm eV} $~(95\%CL,~\planckH+\desi+\DES).
\end{center}
The fact that \DES \, leads to higher neutrino masses compared to \pantheon \, is consistent with the earlier analysis in~\cite{Planck:2018vyg}. \cref{tab:lcdm_bounds} gives a summary of these constraints, including the combination of \planck~ with supernovae data; we note that the addition of supernovae to \planck~ has a similar shift to the replacement of \planck~ by \planckH~ (note also that this shift by adding Pantheon+ has been noticed in~\cite{Wang:2024hen}).

The resulting probability distributions are presented in Fig.~\ref{fig:p18_p20_desi_pantheon_DES}. As one can see, the preferred neutrino masses move to more positive values compared to the \planck+\desi \, case (see~\cref{app:gauss}), which looks promising in view of more precise measurements,  from DESI or Euclid~\cite{Amendola:2016saw}, that could finally confirm a detection of neutrino masses from cosmological data in the region allowed by oscillation experiments $\sum m_\nu>0.06$. The inverted hierarchy scenario is also currently still allowed by the \planckH+\desi+\DES \,  combinations (and only marginally disfavored when considering \planckH+\desi+\pantheon), even with the $\sum m_\nu>0$ prior.

We note that with supernovae data, the addition of more data provides a weaker bound, rather than a stronger one. This is an indication that the datasets in combination here are mildly in tension with respect to the effects of nonzero neutrino mass.

\subsection{Effects of BAO measurements}\label{sec:BAO}

We also investigate here the effect of using other BAO datasets, instead of DESI.
Using the most recent eBOSS DR16 measurements from SDSS~\cite{eBOSS:2020yzd} (in combination with 6DFGS~\cite{Beutler:2011hx} and older data from SDSS) the bound is substantially weaker:
\begin{center}
$\sum m_\nu < 0.14~{\rm eV} $~(95\%CL,~\planckH+\eboss)~. \end{center}
The eBOSS BAO measurements can be also combined with f$\sigma_8$ measurements from redshift-space distortions~\cite{eBOSS:2020yzd}, which has more constraining power:
\begin{center}
$\sum m_\nu < 0.11~{\rm eV} $~(95\%CL,~\planckH+\baoplus)~. \end{center}
Finally, we used the combination of SDSS and DESI as described in \cite{DESI:2024mwx}, leading to
\begin{center}
$\sum m_\nu < 0.11~{\rm eV} $~(95\%CL,~\planckH+\desisdss)~.
\end{center}
This can be considered the state-of-the-art combination, since it uses the measurement with the best BAO statistical power in each redshift bin; this status will shift as DESI releases more data in the future. Note, however, that this combines data processed with different methods/pipelines, and the combination has not been fully validated. We can see here that, with this combination, both inverted and normal hierarchy for the neutrino mass are allowed at 95\% confidence level. 

\subsection{Estimates of Bayesian Evidence for Mass Ordering}

One of the goals of this work has been to evaluate whether the BAO measurements from DESI constitute strong evidence against the inverted mass ordering. To that end, we can compute the Bayes factor (BF), the ratio of the models’ evidences in two models, via the Savage-Dickey ratio. For a model $M_1$ nested within an encompassing model $M_0$, the ratio is given by the ratio of the marginal posterior of $M_0$ evaluated in the limit where it is equivalent to $M_1$ divided by the prior of $M_0$ in the same limit. To evaluate the evidence for or against the inverted mass ordering, let us consider the encompassing model $M_0$ to be the one with the sum of the neutrino masses unconstrained $\sum m_\nu > 0$ and the nested model $M_1$ to have a nonzero lower bound $\sum m_\nu > m_1$. Then, the Bayes factor computed via the Savage-Dickey ratio is written explicitly as
\begin{equation}
    BF_{10} \equiv \frac{P(\mathcal{D}|M_1)}{P(\mathcal{D}|M_0)} = \frac{P(\sum m_\nu > m_1 | \mathcal{D},M_0)}{P(\sum m_\nu>m_1|M_0)}
    \label{SDratio}
\end{equation}
where $BF_{10}$ is the Bayes factor of comparing $M_1$ to $M_0$; $P(\mathcal{D}|M_j)$ is the model evidence, i.e. the probability of the data $\mathcal{D}$ given the model $M_j$; $P(\sum m_\nu > m_1 | \mathcal{D},M_0)$ is the marginalized posterior of the parameter $\sum m_\nu$ integrated over the domain $\sum m_\nu \in [m_1,\infty)$, or the probability of obtaining $\sum m_\nu > m_1$ given $\mathcal{D}$ and $M_0$; and $P(\sum m_\nu>m_1|M_0)$ is the prior on $\sum m_\nu$ for model $M_0$, integrated over the domain $\sum m_\nu \in [m_1,\infty)$. Since we do not bound the domain for the prior on $\sum m_\nu$, we can approximate $P(\sum m_\nu>m_1|M_0) \approx 1$ in the limit where the prior is very broad.\footnote{This approximation introduces an error roughly proportional to the fraction of prior volume contained in the nested model; if the prior boundary was $[0,m_2]$, then the estimated BF would be altered by a factor of $1-m_1/m_2$. Since we evaluate evidence based on $\ln BF$, this error is only $|\ln (1-m_1/m_2)| \approx m_1/m_2 \ll 1$ for $m_1 \ll m_2$.}

Then, if we wish estimate the evidence for or against the inverted mass ordering, we can do this approximately by computing the Bayes factor for a model with a lower bound $m_1=0.1$ eV, which we will denote $BF_{i}$. Note, however, that since this domain also includes normal hierarchy, the Bayes factor may be overestimated (though we do not expect a large change in $\ln BF_i$). The natural logarithms of these Bayes factors are given in \cref{tab:BF}.
Based on, e.g. \cite{Trotta:2008qt}, we can consider $|\ln BF| < 1.0$ to be inconclusive, $1<|\ln BF|<2.5 $ to be ``weak" evidence, $2.5<|\ln BF| <5.0$ to be ``moderate" evidence, and $|\ln BF| > 5.0 $ to be ``strong" evidence. Therefore, we can see that while the combination of \planck+\desi~ provides moderate to strong evidence against the inverted mass ordering, the evidence becomes more moderate with the replacement of \planck~with \planckH, dropping even to the ``weak" evidence category when combined with \DES.

If instead of evaluating the direct evidence for or against inverted mass ordering, one wishes to compare the evidence of one ordering over the other, we could compute the same ratio as in \cref{SDratio}, replacing the encompassing model with the model corresponding to normal ordering $\sum m_\nu > 0.06$ eV. This is equivalent to the ratio of the individual Bayes factors for each model (normal and inverted) with respect to the encompassing model $M_0$ ($\sum m_\nu > 0$). In this approach, we find the Bayes factor for inverted ordering with respect to normal ordering $BF_{i/n}$ and report them in \cref{tab:BF}.
In each case, we find the preference for normal ordering over inverted to be only weak, with the combination \planck+\desi+\DES~nearing on inconclusive.

\begin{table}[t]
    \centering
    \begin{tabular}{|l|c|c|}
    \hline
         Dataset & $\ln BF_{i}$  & $\ln BF_{i/n}$  \\
         \hline
         \planck+\desi & $-4.5$  & $-2.2$ \\
         \planckH+\desi & $-3.7$ & $-1.8$ \\
         \planckH+\desi+\DES & $-2.4$ & $-1.2$ \\
         \hline
    \end{tabular}
    \caption{The natural logarithm of the Bayes factors providing evidence for inverted hierarchy $\ln BF_i$ and evidence for  inverted over normal hierarchy $\ln BF_{i/n}$, computed via the Savage-Dickey ratio, for several datasets.}
    \label{tab:BF}
\end{table}

\section{Constraints on neutrino masses in extensions with Dark Radiation}\label{sec:fluidDR}

Let us first point out the fact that, in $\Lambda$CDM, the sum of the neutrino masses is negatively correlated with $H_0$, as seen in Fig.~\ref{fig:p18_p20_desi_pantheon_DES}. Thus, one may anticipate that a combined analysis with SH$0$ES would drive the fit towards smaller (or even negative, see~\cite{eBOSS:2020yzd,Craig:2024tky}) neutrino masses. This is precisely the case in $\Lambda$CDM, where it is actually inconsistent to combine with SH0ES (+\shoes; see~\cref{app:shoes} for further discussion of this effect). 

On the other hand, in the Dark Radiation (DR) models which alleviate the Hubble tension~\cite{Allali:2024cji}, one may suspect that the same negative correlation exists. However, nonzero neutrino mass exhibits a slight positive correlation with the DR abundance $\DNeff$, defined as the effective number of extra neutrino species $\DNeff\equiv\rho_{\text{DR}}/\rho_\nu$, with $\rho_{\text{DR}}$ and $\rho_\nu$ being the energy densities of DR and one neutrino species, respectively. Since it is known that $H_0$ and $\DNeff$ are positively correlated, the end result for the correlation of $\sum m_\nu$ and $H_0$ is not obvious. 

\begin{table}
    \centering
    \begin{tabular}{|l|c|}
    \hline
         Dataset & $\sum m_\nu$  \\
         \hline
         \planckH+\desi+\pantheon & $< 0.13$\\
         \planckH+\desi+\pantheon+\shoes & $< 0.15$\\
         \planckH+\desi+\DES & $< 0.15$\\
         \planckH+\desisdss+\pantheon & $< 0.15$\\
         \planckH+\desisdss+\DES & $< 0.17$\\
         
         \hline
    \end{tabular}
    \caption{95\%CL upper bounds on the neutrino mass sum for the $\Lambda$CDM + Fluid DR + $\sum m_\nu$ model, comparing different datasets.}
    \label{tab:fluid}
\end{table}

\begin{figure}
    \centering
    \includegraphics[width=0.6\textwidth]{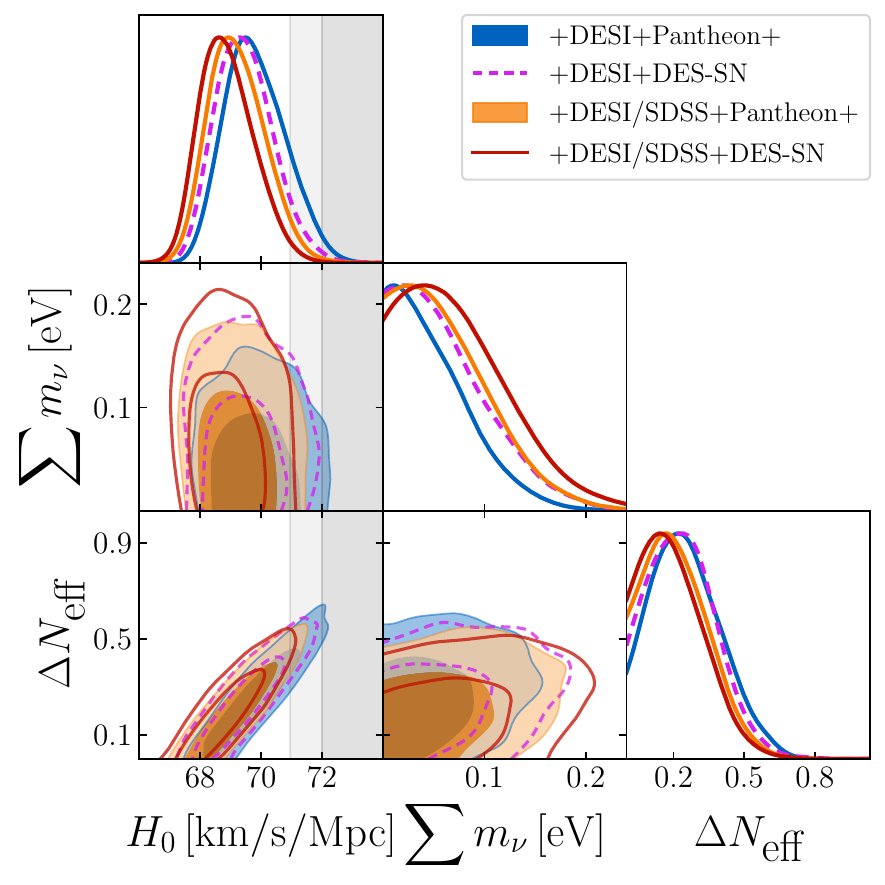}
    \caption{One- and two-dimensional posteriors for $H_0$, $\sum m_\nu$, and $\DNeff$ in the fluid DR model, fitting to combinations of \planckH \, with either \desi \, or \desisdss, and either \pantheon \, or \DES. The inner and outer two-dimensional contours give the 1- and 2-$\sigma$ confidence intervals, respectively. The dark and light gray bands show the 1- and 2-$\sigma$ confidence intervals from the measurement of $H_0$ by the SH0ES collaboration.}
    \label{fig:fluid}
\end{figure}

 We highlight the case of a perfect (self-interacting) fluid DR, over the case where DR is free-streaming, given that the fluid DR relaxes the Hubble tension more significantly~\cite{Allali:2024cji}. We show results in~\cref{tab:fluid} and in~\cref{fig:fluid} for the fluid DR model across several datasets. We focus on the case where the fluid is present during the epoch of big-bang nucleosynthesis (BBN), but we have checked that the conclusions regarding neutrino mass are unaltered for the case where DR is produced after BBN.

Looking first at the constraints on the neutrino mass in~\cref{tab:fluid}, we find overall larger values for the sum of neutrino masses within DR models compared to the $\Lambda$CDM model. Computing Bayes Factors as in the previous section, we find that the evidence against inverted hierarchy is consistently weak in the datasets presented in \cref{tab:fluid}, as low as $\ln BF_{i} \approx -1.2$ for \planckH+\desisdss+\DES. Comparing, instead, with respect to normal ordering, the evidence is inconclusive ($|\ln BF_{i/n} | \lesssim 1$).

Next, it can be appreciated in~\cref{fig:fluid} that the degeneracy between $H_0$ and $\sum m_\nu$ exhibited in $\Lambda$CDM has been broken, and there is no longer a correlation. It is not surprising, therefore, that we see that the addition of the \shoes \, dataset does not substantially alter the neutrino mass bound in~\cref{tab:fluid} (see~\cref{app:shoes} for further comparison). Moreover, we can see in~\cref{fig:fluid} that the posteriors for $\sum m_\nu$ are beginning to form peaks at nonzero values (see~\cref{app:gauss} for a discussion of the peak locations). For example, with the dataset \planckH+\desisdss+\DES, we see a clear peak in the posterior at a value $\sum m_\nu\sim 0.04$ eV, which is $<0.5\sigma$ away from the expected value of $0.06$ eV from neutrino oscillation experiments. Therefore, in the context of the fluid DR model as a solution to the Hubble tension, it is conceivable that a small increase in precision from upcoming data may also lead to the detection of neutrino masses. Note that all of the cases presented in~\cref{fig:fluid} exhibit a $<3\sigma$ tension with SH0ES; however, the case with the highest neutrino masses also exhibits the highest degree of tension with SH0ES, so it remains important to understand this interplay further.

%%%%%%%%%%%%%%%%%%%%%%%%%%%%%%%%%%%%%%%%%%%%%%%%%%%%%%%%%%%%%%%%%%%%%%%%

%\vspace{4mm}
\section{Conclusions}\label{sec:discussions}

\begin{figure}
    \centering
    \includegraphics[width=0.48\textwidth]{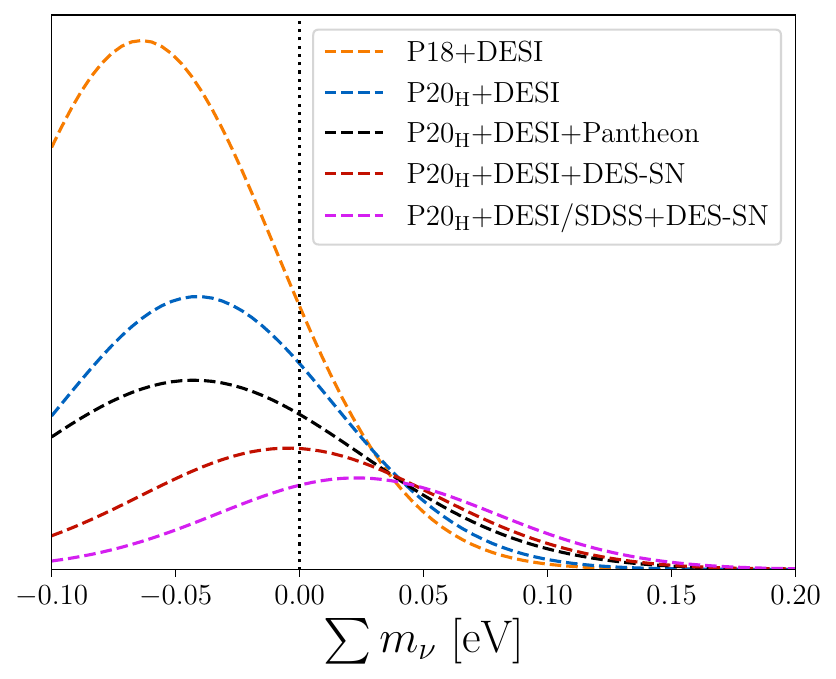}
    \includegraphics[width=0.48\textwidth]{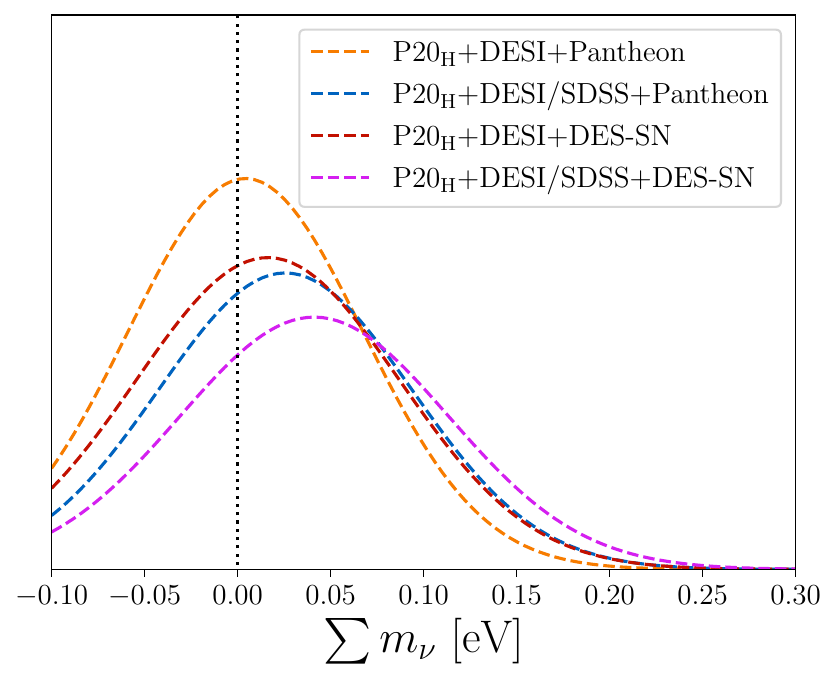}
    \caption{Posterior distributions for $\sum m_\nu$, inferred with the prior $\sum m_\nu>0$, are fit to Gaussian distributions and extended for $\sum m_\nu < 0$ to project the approximate location of the peak preferred by the data. These curves are normalized such that the portion with $\sum m_\nu >0$, corresponding to the posteriors from the MCMC analysis, integrates to unity. The left side figure shows various fits for the $\Lambda$CDM$+\sum m_\nu$ model, while the right side shows the $\Lambda$CDM$+\sum m_\nu$+ Fluid DR model.}
    \label{fig:gauss}
\end{figure}

The new DESI 2024 data release, combined with Planck18 CMB data,  at face value leads to a strong bound on neutrino masses in the $\Lambda$CDM model~\cite{DESI:2024mwx}, when assuming a $\sum m_\nu >0$ prior, which excludes the inverted hierarchy scenario and seems to go even in the direction of negative masses~\cite{eBOSS:2020yzd,Craig:2024tky}. We have shown that these conclusions do {\it not} hold when using more recent  Planck 2020 likelihoods (\hillipop+\lollipop), in combination with cosmological supernovae, since both go in the direction of favoring more positive masses. Concretely, we showed that the Bayesian evidence (quantified by Bayes factors) trends from moderate to weak evidence against inverted hierarchy when including these data, as well as from weak towards inconclusive evidence for normal ordering over inverted. This can be emphasized by extending the posteriors for $\sum m_\nu$ to negative values by fitting to a Gaussian; \cref{fig:gauss} shows that the peaks of the would-be posteriors are shifted towards positive values by adding more data, especially in the case of DES supernovae (see \cref{app:gauss} for more details).
The distributions are peaked at even larger positive values of $\sum m_\nu$ for the fluid DR model, closer to the expected lower bound of $\sum m_\nu > 0.06$ eV from neutrino oscillations.
We find the conservative upper bound at 95$\%$CL to be $m_\nu<0.12$ eV ($m_\nu<0.1$ eV) when adding also DES-SN5YR (or Pantheon$+$) supernovae. 

Furthermore, we have analyzed neutrino mass bounds in a model with fluid Dark Radiation that addresses the Hubble tension~\cite{Allali:2024cji}, and we have shown that in this case: (i) neutrino mass bounds are driven to even larger values, (ii) bounds are robust when combining with the SH$0$ES measurement of $H_0$, and (iii) posterior probabilities even peak at nonzero neutrino masses.

Even if our findings go in the direction of relaxing constraints, they in fact constitute a significant improvement in the consistency with the expectation of $\sum m_\nu \geq 0.06$ eV that comes from neutrino oscillation experiments. Overall, our results represent a promising starting point in the quest for neutrino mass detection with upcoming cosmological data.

\begin{acknowledgments}
We thank Fabrizio Rompineve and Marko Simonovic for useful discussions. The work of A.N. is supported by the grants PID2019-108122GB-C32 from the Spanish Ministry of Science and Innovation, Unit of Excellence Maria de Maeztu 2020-2023 of ICCUB (CEX2019-000918-M) and AGAUR 2021 SGR 00872. The work of I.J.A. is supported by NASA grant 80NSSC22K081. A.N. is grateful to the Physics Department of the University of Florence for the hospitality during the course of this work. Part of this work was conducted using computational resources and services at the Center for Computation and Visualization, Brown University.
We also acknowledge use of the INFN Florence cluster.
\end{acknowledgments}

%%%%%%%%%%%%%%%%%%%%%%%%%%%%%%%%%%%%%%%%%%
\bibliographystyle{JHEP}
\bibliography{biblio.bib}

\appendix 

\begin{center}
\Large{Appendix}
\end{center}

\section{Combined analysis with SH0ES}\label{app:shoes}

Let us explore the effect of combining with SH0ES on neutrino masses.

\begin{figure}
    \centering
    \includegraphics[width=0.6\textwidth]{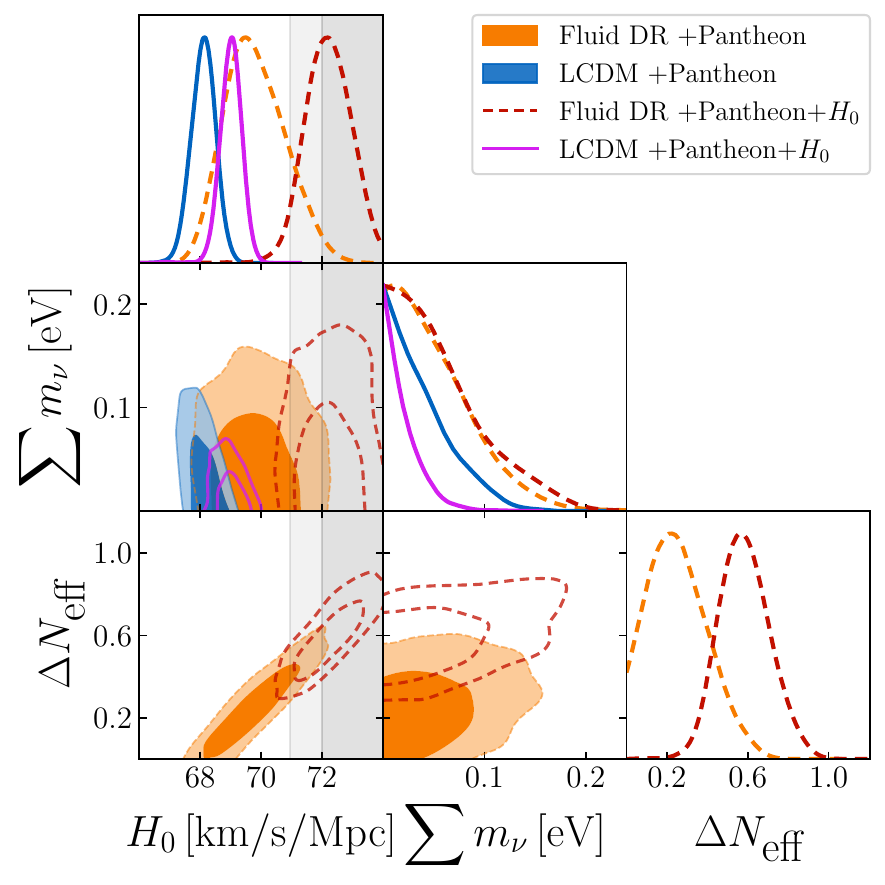}
    \caption{One- and two-dimensional posteriors for $H_0$, $\sum m_\nu$, and $\DNeff$ in both $\Lambda$CDM and the fluid DR model, fitting to \planckH+\desi+\pantheon~ and the same set with the addition of \shoes. The inner and outer two-dimensional contours give the 1- and 2-$\sigma$ confidence intervals, respectively. The dark and light gray bands show the 1- and 2-$\sigma$ confidence intervals from the measurement of $H_0$ by the SH0ES collaboration.}
    \label{fig:shoes}
\end{figure}

First, in the context of $\Lambda$CDM, we can see in~\cref{fig:shoes} that, due to the degeneracy between $H_0$ and $\sum m_\nu$, a combination with SH0ES would cause a dramatically tighter constraint on $\sum m_\nu$. In this case, the would-be constraint from the combination \planckH+\desi+\pantheon\\
+\shoes~ is $\sum m_\nu < 0.055 $ eV, putting this constraint in conflict with neutrino oscillation experiments. Note that that the datasets in this combination are in great tension and therefore this combination is not  justified. 

Instead, in the case of fluid DR, the degeneracy between $H_0$ and $\sum m_\nu$ is no longer present. As seen in~\cref{tab:fluid}, this means that the constraint on $\sum m_\nu$ does not become tighter when adding SH0ES (now justified due to the lesser tension); in fact, it becomes even a bit weaker (shifting from $<0.13$ eV with \planckH+\desi+\pantheon~ to $<0.15$ eV when adding +\shoes).~\cref{fig:shoes} exhibits this as well, seen in the fact that the one-dimensional posterior for $\sum m_\nu$ is not made tighter by SH0ES.

\section{Neutrino mass posteriors fit to a Gaussian}\label{app:gauss}

We discuss now an assessment of whether the neutrino mass posteriors are peaked at would-be negative values of the neutrino mass, as was considered in~\cite{eBOSS:2020yzd,Craig:2024tky}. To do so, we can take the posteriors which are inferred when using a prior of $\sum m_\nu >0$, as we have done in this work, and fit them to the tail of a Gaussian. Then, one can project where the preferred peak would lie if the fit were to extend to negative masses. Note that in~\cite{Craig:2024tky}, a different method is proposed.

We see in~\cref{fig:gauss} that for some datasets, the $\Lambda$CDM$+\sum m_\nu$ model exhibits distributions that would peak at negative values. However, the inclusion of supernovae data drives these peaks closer to zero, mostly due to the DES-SN5YR supernovae. Also, when combining the DESI BAO and SDSS BAO measurements along with DES-SN5YR supernovae, we can see even a peak at positive values. Further, using the same technique underscores the fact that in the $\Lambda$CDM$+\sum m_\nu$+ Fluid DR model, the peaks are definitively driven to positive values of $\sum m_\nu$.

\section{Constraints from various BAO measurements}

\begin{figure}
    \centering
    \includegraphics[width=0.5\textwidth]{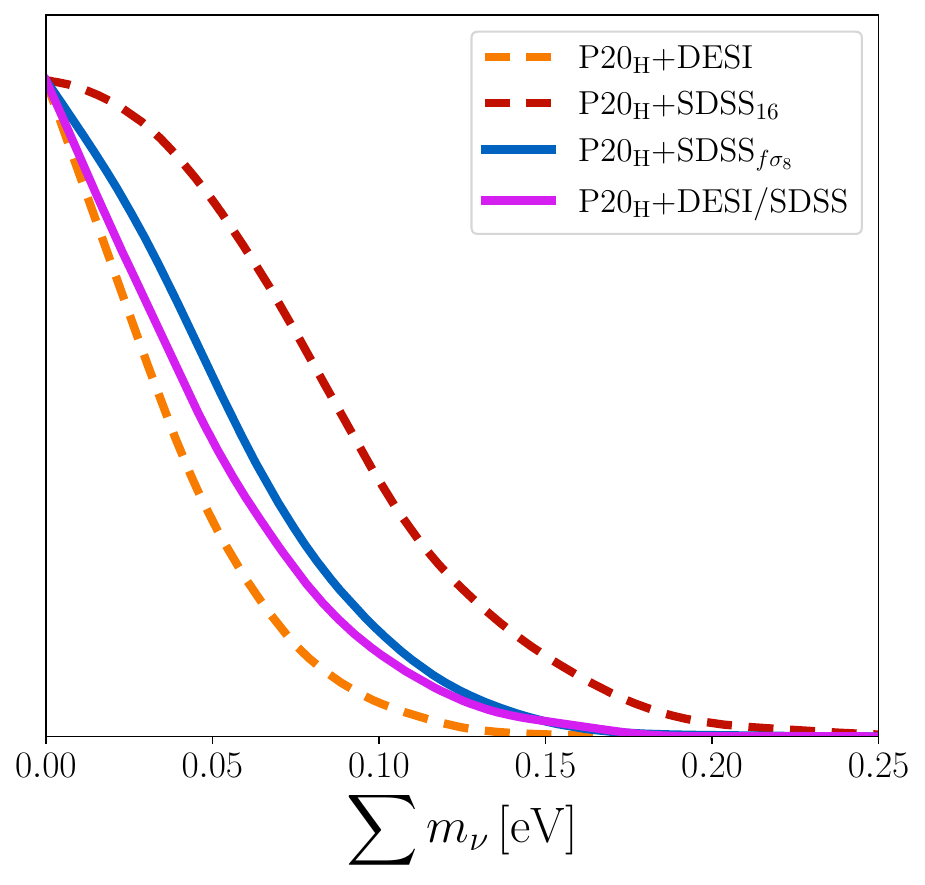}
    \caption{One-dimensional posteriors for $\sum m_\nu$, fitting to combinations of \planckH~with different BAO datasets: \desi,\, \eboss, \baoplus, or \desisdss.}
    \label{fig:p20_bao}
\end{figure}

\cref{fig:p20_bao} gives the posteriors for $\sum m_\nu$ comparing the effects of different BAO datasets, as discussed in \cref{sec:BAO}.

\end{document}